\documentclass[%
prc,
reprint,
superscriptaddress,
showpacs,
showkeys,
nofootinbib,
 amsmath,amssymb,
 aps,
]{revtex4-1}

\usepackage{graphicx}
\usepackage[T1]{fontenc} 
\usepackage{epstopdf} 
\usepackage{dcolumn}
\usepackage{bm}
\usepackage{hyperref}
\usepackage[mathlines]{lineno}
\usepackage{xfrac} 
\usepackage{color}

\begin{document}
\preprint{APS/123-QED}

\title{Shape coexistence in the neutron-deficient lead region: A systematic study of lifetimes in the even-even $^{188-200}$Hg with GRIFFIN}


\author {B.~Olaizola}
\email{e-mail: bruno.olaizola@triumf.ca}
\affiliation{TRIUMF, 4004 Wesbrook Mall, Vancouver, British Columbia V6T 2A3, Canada.}

\author {A.B.~Garnsworthy}
\affiliation{TRIUMF, 4004 Wesbrook Mall, Vancouver, British Columbia V6T 2A3, Canada.}

\author {F.A.~Ali}
\affiliation{Department of Physics, University of Guelph, Guelph, Ontario N1G 2W1, Canada.}
\affiliation{Department of Physics, College of Education, University of Sulaimani, P.O. Box 334, Sulaimani, Kurdistan Region, Iraq}

\author {C.~Andreoiu}
\affiliation{Department of Chemistry, Simon Fraser University, Burnaby, BC V5A 1S6, Canada.}

\author {G.C.~Ball}
\affiliation{TRIUMF, 4004 Wesbrook Mall, Vancouver, British Columbia V6T 2A3, Canada.}

\author {N.~Bernier}
\affiliation{TRIUMF, 4004 Wesbrook Mall, Vancouver, British Columbia V6T 2A3, Canada.}
\affiliation{Department of Physics and Astronomy, University of British Columbia, Vancouver, BC, V6T 1Z4, Canada}

\author {H.~Bidaman}
\affiliation{Department of Physics, University of Guelph, Guelph, Ontario N1G 2W1, Canada.}

\author {V.~Bildstein}
\affiliation{Department of Physics, University of Guelph, Guelph, Ontario N1G 2W1, Canada.}

\author {M.~Bowry}
\affiliation{TRIUMF, 4004 Wesbrook Mall, Vancouver, British Columbia V6T 2A3, Canada.}

\author {R.~Caballero-Folch}
\affiliation{TRIUMF, 4004 Wesbrook Mall, Vancouver, British Columbia V6T 2A3, Canada.}

\author {I.~Dillmann}
\affiliation{TRIUMF, 4004 Wesbrook Mall, Vancouver, British Columbia V6T 2A3, Canada.}
\affiliation{Department of Physics and Astronomy, University of Victoria, Victoria, British Columbia V8P 5C2, Canada}

\author {G.~Hackman}
\affiliation{TRIUMF, 4004 Wesbrook Mall, Vancouver, British Columbia V6T 2A3, Canada.}

\author {P.E.~Garrett}
\affiliation{Department of Physics, University of Guelph, Guelph, Ontario N1G 2W1, Canada.}

\author {B.~Jigmeddorj}
\affiliation{Department of Physics, University of Guelph, Guelph, Ontario N1G 2W1, Canada.}

\author {A.I.~Kilic}
\altaffiliation{Present address: Nuclear Physics Institute of ASCR, 250 68 \u{R}e\u{z}. Prague, Czech Republic}
\affiliation{Department of Physics, University of Guelph, Guelph, Ontario N1G 2W1, Canada.}

\author {A.D.~MacLean}
\affiliation{Department of Physics, University of Guelph, Guelph, Ontario N1G 2W1, Canada.}

\author {H.P.~Patel}
\affiliation{TRIUMF, 4004 Wesbrook Mall, Vancouver, British Columbia V6T 2A3, Canada.}

\author {Y.~Saito}
\affiliation{TRIUMF, 4004 Wesbrook Mall, Vancouver, British Columbia V6T 2A3, Canada.}
\affiliation{Department of Physics and Astronomy, University of British Columbia, Vancouver, BC, V6T 1Z4, Canada}

\author {J.~Smallcombe}
\altaffiliation{Present Address: Oliver Lodge Laboratory, The University of Liverpool, Liverpool, L69 7ZE, UK}
\affiliation{TRIUMF, 4004 Wesbrook Mall, Vancouver, British Columbia V6T 2A3, Canada.}

\author {C.E.~Svensson}
\affiliation{Department of Physics, University of Guelph, Guelph, Ontario N1G 2W1, Canada.}

\author {J.~Turko}
\affiliation{Department of Physics, University of Guelph, Guelph, Ontario N1G 2W1, Canada.}

\author {K.~Whitmore}
\affiliation{Department of Chemistry, Simon Fraser University, Burnaby, BC V5A 1S6, Canada.}

\author {T.~Zidar}
\affiliation{Department of Physics, University of Guelph, Guelph, Ontario N1G 2W1, Canada.}

\date{\today}

\begin{abstract}
Lifetimes of $2^+_1$ and $4^+_1$ states, as well as some negative-parity and non-yrast states, in $^{188-200}$Hg were measured using $\gamma-\gamma$ electronic fast timing techniques with the LaBr$_3$(Ce) detector array of the GRIFFIN spectrometer. The excited states were populated in the $\epsilon/\beta^+$-decay of $J^\pi =7^+/2^-$ $^{188-200}$Tl produced at the TRIUMF-ISAC facility. The deduced B(E2) values are compared to different interacting boson model predictions. The precision achieved in this work over previous ones allows for a meaningful comparison with the different theoretical models of these transitional Hg isotopes, which confirms the onset of state mixing in $^{190}$Hg.

\end{abstract}

\pacs{
21.10.-k, 
21.10.Tg, 
23.20.-g, 
23.20.Lv, 
27.70.+q,	
27.80.+w	
}
\keywords{$^{188}$Hg, $^{190}$Hg, $^{192}$Hg, $^{194}$Hg, $^{196}$Hg, $^{198}$Hg, $^{200}$Hg, fast timing, LaBr$_3$(Ce), B(E2)}

\maketitle


\section{Introduction}

Shape coexistence is a unique phenomenon of the atomic core in which the nucleus displays intrinsically different shapes within a small energy range. Manifestation of this behaviour has been observed all across the nuclear chart, but the neutron-deficient Pb region (\textit{Z}$\leq$82, $N$<126) is characterized by some of the clearest examples of shape coexistence~\cite{HEY11,AND00,Dracoulis2003,Dracoulis2004,Julin2016,Marsh2018}. The phenomenon was observed in the Pb isotopes using $\alpha$-decay spectroscopy, which found multiple low-lying $0^+$ states in $^{186}$Pb \cite{AND00}. It was shown by Dracoulis \textit{et al.}~\cite{DRA00}, that the high-spin isomeric states in $^{188}$Pb can only be built on unique single-particle configurations of different shape. This clearly demonstrated that three differently shaped potentials (spherical, prolate and oblate) exist in these nuclei.

In the light Hg ($Z$=80) isotopes, this phenomenon was first revealed in optical spectroscopy measurements which identified a large staggering in the isotope shifts between the odd and even Hg isotopes~\cite{BON72}. This isotope shift was interpreted as an alternation between normal and intruder configurations being the ground state with the removal of neutrons. Later laser spectroscopy studies have determined that $^{181}$Hg represents the lighter end of the staggering, and also confirmed the inversion between the ground state and isomeric state in $A$=185~\cite{Marsh2018, Sels2019}. In the even Hg, only recently, a Coulomb excitation study obtained detailed spectroscopic information about shape coexistence for  $^{182-188}$Hg~\cite{Bree2014}. By measuring the relative sign of the \textit{E2} matrix elements, Bree \textit{et al.}~\cite{Bree2014} were able to extract information about the different deformations of the $0^+$ states and firmly establish that two different structures coexist at low energies. 

Despite these ground-breaking experiments, there is still a significant amount of key information that remains to be measured, especially in the transitional isotopes between the stable $^{200}$Hg and the beginning of the midshell $^{190}$Hg. This experimental data is critical for solidifying our understanding of the region and developing a quantitative understanding of the underlying mechanisms driving these behaviours. The relative energy of the intruder states has a parabola-shape with a minimum observed at $^{182}$Hg. In the heavier transitional isotopes ($190\leq A\leq 200$), the ground and intruder configurations are still sufficiently far apart in energy such that the mixings between the two structures are expected to be significantly reduced. These isotopes thus present a good opportunity to benchmark the normal ground-state configuration without the perturbations (through mixing with the intruder configuration) experienced in the lighter isotopes, thus simplifying the comparison with different theoretical calculations.

One of the main model-independent probes used to study shape coexistence is the measurement of transition strengths, in particular $B(E2)$ and $\rho^2$($E0$) values~\cite{HEY11}. These transition strengths are particularly sensitive to the wavefunctions of the states they connect, and thus are one of the most stringent probes available to test theoretical models used to describe nuclei. With respect to $B(E2)$ values in the transitional Hg, Esmaylzadeh \textit{et al.}~\cite{Esmaylzadeh2018} recently measured the $2^+_1$ lifetimes for $^{190,192,194}$Hg. Due to the isotope production mechanism employed, the experiment suffered from contaminants that significantly limited the precision of the measured half-lives, preventing a meaningful comparison with different theoretical calculations. In the case of $\rho^2$($E0$) values, the excited $0^+$ states have only been identified up to $^{190}$Hg, with the energies of the intruder structures remaining unknown for the heavier isotopes, although some candidate states exist. Theoretical calculations predict an increase in excitation energy for the intruder configuration up to $^{192}$Hg after which a more stable value is maintained~\cite{Nomura2013, Garcia-Ramos2014}.

In order to characterize the evolution from the stable Hg isotopes towards the mid-shell, a systematic study of the decay of the ground- and isomeric states of neutron-deficient $^{188-200}$Tl isotopes into Hg has been performed using the GRIFFIN spectrometer~\cite{GAR19,SVE14,RIZ16} at TRIUMF-ISAC. The high statistics resulting from the measurement of $\gamma$-ray and conversion electrons enable high precision $\gamma-\gamma$ angular correlations and precise branching ratios, which are all important in forming a complete picture of the band structure of these isotopes. In the present article, we focus on the results of the lifetime measurements. Data collected with the ancillary LaBr$_3$(Ce) array have been analyzed using the Generalized Centroid Difference Method (GCDM)~\cite{Regis2013} to precisely measure lifetimes of all the first 2$^+$ and 4$^+$ states of the ground-state bands, as well as some negative-parity and non-yrast states. The extracted $B(E2)$ values are compared with different interacting boson model (IBM) calculations while the negative-parity band is interpreted in comparison with a quasiparticle-rotor model.




\section{Experimental setup\label{sec:experimental_setup}}

 The Tl isotopes were produced by a 500-MeV proton beam of 9.8~$\mu$A intensity delivered by the TRIUMF main cyclotron \cite{BYL14} impinging on an uranium carbide (UC$_x$) target. The TRIUMF Resonant Ionization Laser Ion Source (TRILIS)~\cite{LAS06} was tuned to preferentially ionize the 7$^+$ isomeric states in Tl. A small contribution of the $2^-$ ground state Tl was also present in the beam, but no other significant isobaric contaminants were observed. The Tl ion beam was accelerated to 20~kV, mass separated and delivered to the experimental station. The beam intensity was attenuated down to $\sim 10^5$ particles per second for all the masses studied.
 
 The ions were implanted in a mylar tape at the central focus of the GRIFFIN spectrometer~\cite{GAR19,SVE14,RIZ16}. Due to the long half-lives involved in the Hg decay chains and the low-energy $\gamma$-ray transitions of the Hg decay products (all below the energy range of interest in this experiment), the tape remained stationary during the beam delivery. The tape was moved only when changing between beams of different mass in order to remove any remaining longer-lived activity from the previous setting from the chamber. The exception was for the decay of $^{188m}$Tl, where the tape cycling mode was used. The cycle was composed of 1.5~s for the tape movement, 30~s of background measurement, 480~s of the beam being delivered and just 1~s of decay time with the beam blocked. This cycling configuration was designed to maximize the activity of $^{188m}$Tl while suppressing the other decay products present. In order to remeasure the $^{188}$Tl and $^{188m}$Tl half-lives (T$_{1/2}(2^-)$=71(1)~s and T$_{1/2}(7^+)$=71(2)~s, respectively~\cite{Kondev2018}), a small fraction of the data was taken with a different tape cycle; 1.5~s of tape movement, 30~s background measurement, 210~s beam-on and 350~s of decay time with the beam off. The detailed analysis of the decays from these two levels is being prepared for publication \cite{MacLean2020}.

GRIFFIN is an array of up to 16 high-purity germanium (HPGe) clover detectors~\cite{RIZ16} arranged in a rhombicubocatahedral geometry. For this particular experiment, only 15 HPGe-clovers were employed as one must be removed to accommodate the liquid nitrogen dewar of the PACES detector. Seven cylindrical (5.1~cm in diameter by 5.1~cm length) lanthanum bromide crystals doped with a $5\%$ of cerium (LaBr$_3$(Ce)) coupled to a R2083 photomultiplier (PMT) were placed in the ancillary triangular positions of the array (one ancillary position remained empty). Around the implantation point, covering the upstream half of the chamber, a set of five in-vacuum LN$_2$-cooled lithium-drifted silicon detectors (PACES) were used for conversion electron measurements. A fast 1\,mm-thin plastic called Zero Degree Scintillator (ZDS) was placed just a few millimeters behind the ion-deposition point in the tape. The reader is referred to~\cite{GAR19} for further details about the GRIFFIN array and ancillary detector performance.

The energy signals from each detector were digitazed by the GRIFFIN custom-built digital data acquisition system (DAQ)~\cite{GAR17} with a 100~MHz sampling frequency, which, after a digital implementation of a constant-fraction discriminator (CFD) algorithmic interpolation, gives timestamps with a precision down to $\sim 1$~ns. After shaping the signals with a custom made preamplifier, this works well for the timestamps of the HPGe and Si(Li) semiconductor signals, as well as the signals from the bismuth germanate (BGO) shields and SCEPTAR, the thick $\beta$-tagging plastic scintillators (neither of them employed in this experiment). However, it is not sufficient for accurate timing of the fast ZDS and LaBr$_3$(Ce) signals which have a rise time of 0.7~ns Ref.~\cite{Vedia2015}. To make use of the full timing capabilities of these fast scintillators, a hybrid analog-digital electronic timing setup was developed and employed in the present work. 

The LaBr$_3$(Ce) energy signal is taken from the last dynode of the PMT and processed by a custom-made preamplifier before being directly provided to the DAQ. The timing signal is taken from the anode and input to an Ortec 935 quad CFD~\cite{ortec935}. External delay cables of 20~ns were employed in order to obtain an uniform time walk over a large energy range while maintaining reasonable timing resolutions. The output of the individual CFDs were fed into Lecroy 429A fan-in/fan-out logic modules~\cite{lecroy429A} in order to obtain all the possible LaBr$_3$(Ce)-LaBr$_3$(Ce) combinations. These logic output signals provide the START and STOP signals to a set of Ortec 566 Time-to-Amplitude Converter (TAC) modules~\cite{ortec566}. These TAC modules delay the output signal by $\sim 2.5 \mu$s which is then digitized by the same 100 MHz ADC described above. . The timestamps were corrected offline to time-match the detector and TAC signals.

Simultaneously the signals from an Ortec 462 Time Calibrator module~\cite{ortec462} operated at a low rate of $\sim 100$~Hz were connected to the TAC modules during the whole data-collecting period. The Time Calibrator has a timing precision of $\sim 10$~ps. This allowed for a precise monitoring of the TAC performance and corrections to any fluctuations due to, for example, temperature changes. These events were easily identified through a lack of LaBr$_3$(Ce) energy coincidence. An offline event-by-event correction of the TACs was performed using this information.

In this configuration the combination of all seven LaBr$_3$(Ce) crystals has a timing resolution of $\sim 330$~ps, with a time-walk of slightly over 100~ps in the 200-1300 keV range, as shown in Fig.~\ref{fig:prd}. Further details on the data shown in this Figure is provided in the following Section. 

An additional TAC was set with the ZDS plastic as START and a logic module with an OR of all the LaBr$_3$(Ce) detectors as STOP. Thanks to its reduced thickness of 1~mm, it ensures that charged particles will deposit an approximately constant amount of energy nearly independent of their kinetic energy. This allows for a superior timing resolution and a reduced time-walk when compared to LaBr$_3$(Ce). This comes, however, at the cost of losing all $\beta$-particle energy resolution. The ZDS has an absolute efficiency of $\sim 20\%$, due to its solid angle coverage, which is an order of magnitude higher than the LaBr$_3$(Ce) array.

To reduce the volume and rate of data recorded to disk, the DAQ system employed digital filters. Such events with at least one PACES or one HPGe or two or more LaBr$_3$(Ce) crystals had signals were passed to the data acquisition computer. Any other detector hits that were in temporal coincidence within $2~\mu$s of any one of these conditions were also recorded to disk.

\section{Data analysis\label{sec:data_analysis}}

Data collected with the GRIFFIN array were analyzed using the GRSISort software~\cite{GRSISort} within the ROOT framework~\cite{Brun1997}. General methods for analyzing such experiments are outlined in Ref.~\cite{GAR19}. This experiment focused on measuring lifetimes in the pico- to nanosecond range using the GCDM. This method is an evolution of early electronic fast-timing techniques~\cite{Mach1989,Moszynski1989}, adapted to large arrays of fast inorganic scintillators~\cite{Regis2010,Regis2013}. A detailed explanation of the GCDM can be found in Refs.~\cite{Regis2010,Regis2012,Regis2013,Regis2014,Regis2016}, but a short summary of the method is included here.

\subsection{Centroid difference}

\begin{figure}
\begin{center}
\includegraphics[width=\columnwidth, keepaspectratio]{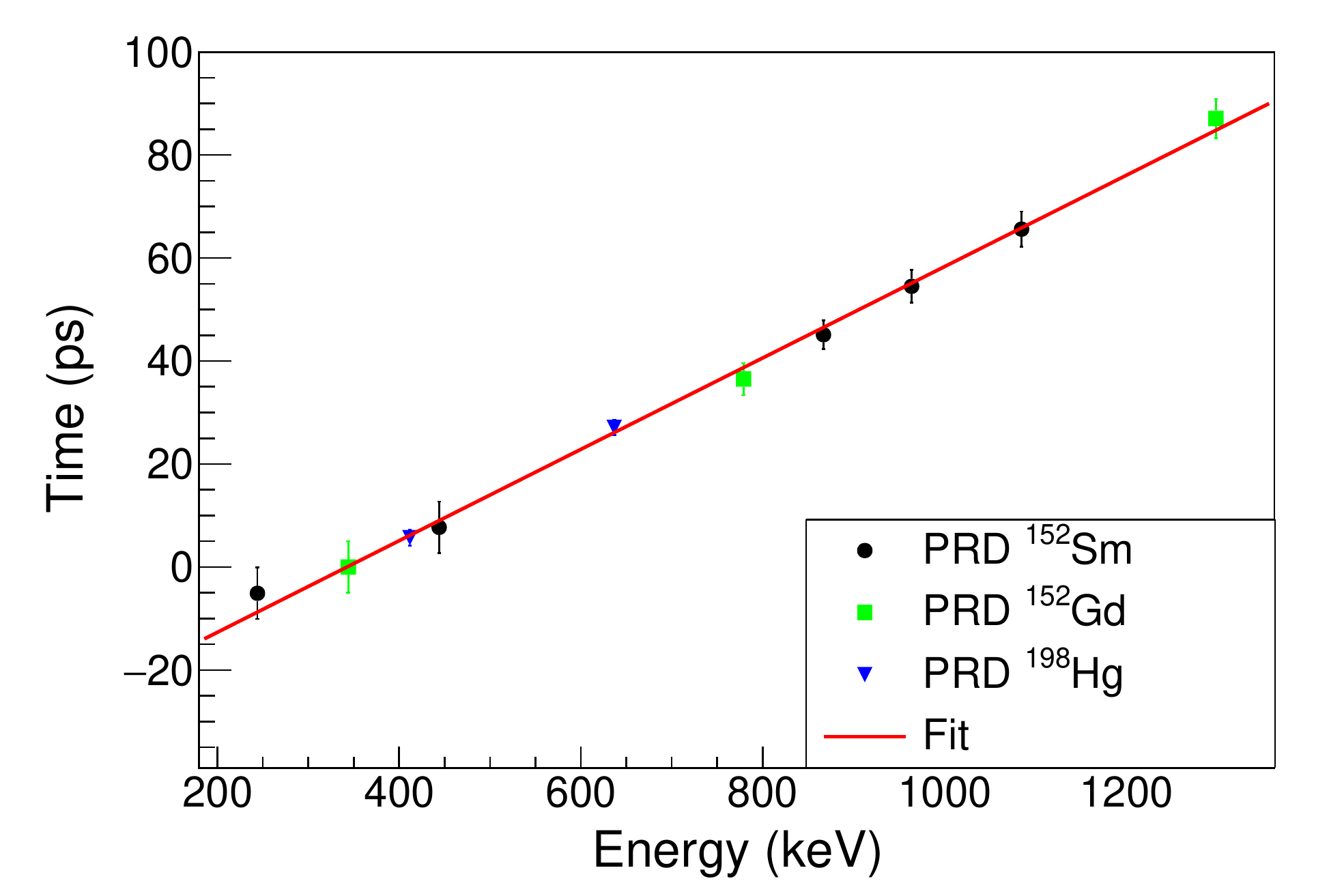}
\caption{\label{fig:prd} Time-walk response or PRD curve of the whole LaBr$_3$(Ce) array. The curve has been derived from data collected with an offline commercial $^{152}$Eu source and with $^{198m}$Tl online. See text for details.}
\end{center}
\end{figure}

The method uses a TAC to measure the time difference between two transitions in a $\gamma$-ray cascade detected by LaBr$_3$(Ce) crystals. If the decaying $\gamma$-ray is prompt (that is, if the draining transition decays from a nuclear state with a lifetime well below the timing sensitivity of the system, i.e. $\tau < 1$~ps), the TAC spectrum will be a semi-Gaussian distribution centered at zero. When the intermediate level between the two transitions has a mean lifetime in the picosecond or longer range, the resulting distribution centroid will be shifted from zero by an amount equal to $\tau$. Despite the use of analog CFDs in the signal processing, the position of the \textit{prompt} zero-time will depend on the energy of the feeding and decaying transitions. This is known as the time-walk or ``mean prompt response difference'' (PRD(E)) for arrays, and can be calibrated down to 2-5~ps using standard commercially-available radioactive sources such as $^{152}$Eu (all the time distribution centroids must be corrected by the precisely known lifetimes). The curve generated from data collected for the present study is shown in Fig.~\ref{fig:prd}.

The cables providing signals to the STOP input of the TAC modules are $\sim 25$~ns longer than the ones feeding the START. In practice, this has the effect of shifting the TAC range from between 0 and 50~ns to between $-25$ and 25~ns, allowing for reverse gating of the transitions in the LaBr$_3$(Ce) crystals. In this anti-delayed mode, the decaying transition is the START input for the TAC and the feeding transition is the STOP. The time difference between the signals will therefore be the negative lifetime of the intermediate level. The centroid difference between the direct and reverse gating is equal to twice the lifetime and with this method most systematic errors are suppressed. The correction for the time walk described earlier must still be included and the final expression is of the form:

\begin{equation}
	\Delta\text{C} =  2\tau + \text{PRD}(\Delta \text{E}_\gamma) \label{eq:centroid_shift}
\end{equation}

\noindent where $\Delta\text{C}$ is the centroid difference between the two gated spectra, $\tau$ is the sum of all the lifetimes of levels between the two gated transitions, and $\text{PRD}(\Delta \text{E}_\gamma) = \text{PRD}(\text{E}_\text{feed}) - \text{PRD}(\text{E}_\text{decay})$ is the difference in the time response for the energies of the feeding and decaying transitions (see Fig.~\ref{fig:prd}).

\begin{figure}
\begin{center}
\includegraphics[width=\columnwidth, keepaspectratio]{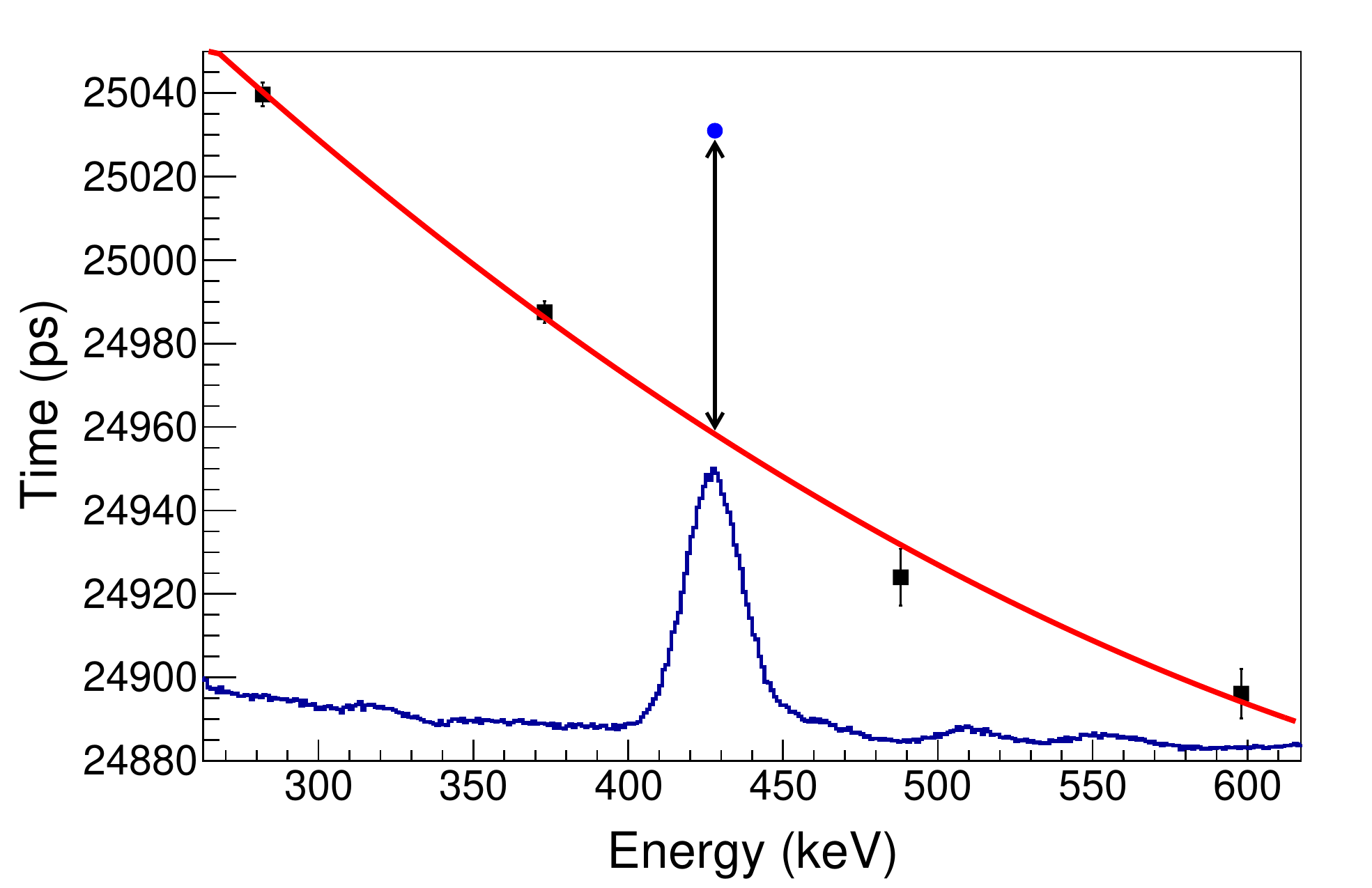}
\caption{\label{fig:compton_corr} Compton correction to the $2^+_1$ mean lifetime in $^{194}$Hg ($E_{\gamma}=428$keV). The black squares are the centroid values of the time response for the Compton background at that energy, the blue circle is the centroid of the time response for that FEP and the red line is the quadratic fit to the Compton time-response. The LaBr$_3$(Ce) energy spectrum is superimposed for reference. All spectra were generated with gates in the $4^+_1 \rightarrow 2^+_1$ transition in the START LaBr$_3$(Ce) and $6^+_1 \rightarrow 4^+_1$ in the HPGe array. See text for details.}
\end{center}
\end{figure}

For the extraction of nuclear lifetimes, the data were sorted into LaBr$_3$(Ce)-LaBr$_3$(Ce)-TAC-(HPGe) events where the detector name implies the energy from that detector. (The HPGe coincidence was optional but allowed to use the GCDM imposing an additional condition on the high-energy-resolution HPGe detectors to precisely select one specific $\gamma$-ray cascade, if needed). This is especially important because of the very different timing response of the LaBr$_3$(Ce) crystals to full-energy peaks (FEP) and Compton events. This difference in timing response makes it impossible to subtract the nearby background when gating on a peak (as it is done in HPGe-HPGe coincidences, for example), since the difference in energies and type of physics event will yield very different time responses. It is, thus, of paramount importance to reduce the background by other means, like imposing additional coincidence conditions or the use of anti-Compton and background shields. Nevertheless, the time response of the Compton background around the peak is studied and a correction is applied using the following equation:

\begin{equation}
    \textrm{A}_\textrm{T} \cdot \textrm{C}_\textrm{T} = \textrm{A}_\textrm{FEP} \cdot \textrm{C}_\textrm{FEP} + \textrm{A}_\textrm{C} \cdot \textrm{C}_\textrm{C} \label{eq:compton_corr}
\end{equation}

\noindent where A stands for area and C for timing-response centroid value. The subscripts refer to full-energy peak ($_\textrm{FEP}$), Compton ($_\textrm{C}$) and the total area of FEP plus Compton ($_\textrm{T}$). The Compton gate is set a few keV above the FEP energy, so, due to the CFD time-walk, C$_\textrm{C}$ must be shifted as a function of energy. Several gates are set on the Compton background around the peak and their centroid values are fitted as a function of the energy. This is shown in Fig~\ref{fig:compton_corr}. For further details on this approach to Compton correction see Refs.~\cite{Mach1989, Regis2013}.

The GRIFFIN array and its ancillary detectors (especially the LaBr$_3$(Ce) array) have a very compact geometry, which causes a significant Compton background. This has been greatly mitigated by the recent addition of BGO active Compton and background suppression shields~\cite{GAR19}. However, since this shielding was not available at the time of the present experiment, an additional coincidence can be set in the HPGe detectors when examining a cascade involving three or more $\gamma$ rays. Alternatively, if the cascade involves only two $\gamma$ rays, anti-coincidence conditions can be imposed with the HPGe data. When $\gamma$ rays involved in cascades of only two $\gamma$ rays are selected in the LaBr$_3$(Ce), it can be assumed that any events in the HPGe detectors will be a Compton or random-background event, and the entire GRIFFIN array can effectively be used as an active suppression shield. This drastically reduced the Compton background, with a minor loss in statistics. During the offline timing calibration, no HPGe data were recorded. This resulted in a peak-to-background ratio in the $^{152}$Eu decay spectra which was much poorer than that achieved in the online data (when the HPGe were active) and in calibrations of subsequent experiments. This has significantly increased the uncertainty in the PRD curve available for this experiment up to $\sim 5$~ps.

\begin{figure}
\begin{center}
\includegraphics[width=\columnwidth, keepaspectratio]{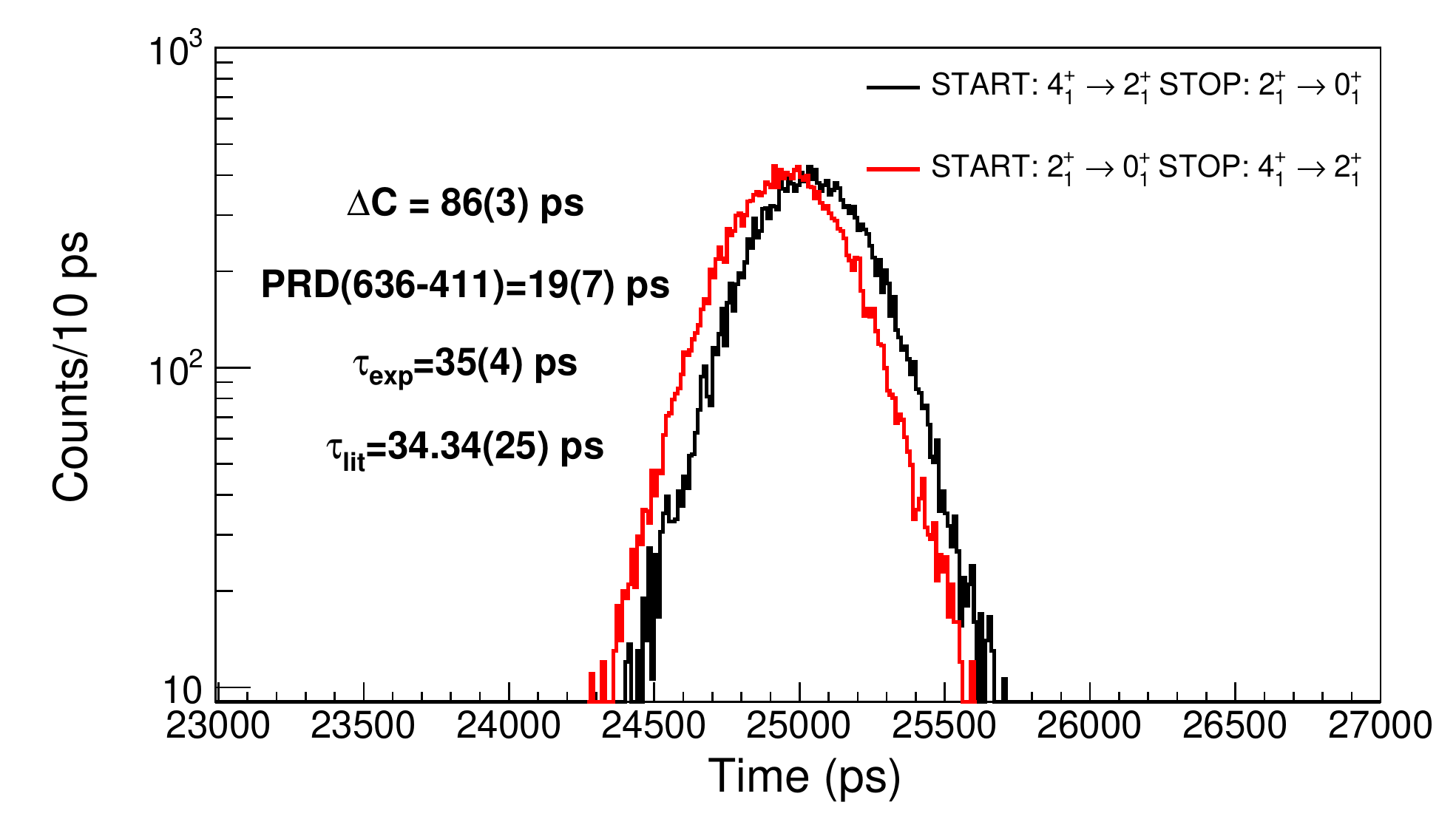}
\caption{\label{fig:Hg198_2+_lifetime} Centroid shift ($\Delta$C) between the delayed (\textit{black}) and anti-delayed (\textit{red/gray}) time spectra of the $4^+_1 \rightarrow 2^+_1$ and $2^+_1 \rightarrow 0^+_1$ $\gamma$-ray transitions in $^{198}$Hg. The $\Delta$C value must be corrected by the PRD(E) and Compton contributions. Only the PRD(E) corrections derived from the $^{152}$Eu source data were applied here (the precise $^{198}$Hg points were excluded) in the determination of this $\tau(2^+_1)$ value, resulting in a comparatively larger uncertainty than for other masses.}
\end{center}
\end{figure}

To improve the precision of the PRD(E) in the energy range of interest (400 to 650~keV), the $2^+_1\rightarrow0^+_1$ and $4^+_1\rightarrow2^+_1$ transitions from $^{198}$Hg were included in the PRD(E) calibration. The lifetime of the $2^+_1$ state in $^{198}$Hg has been measured in over 10 different experiments using a wide range of techniques, yielding a very accurate evaluated value of $\tau=34.34(25)$~ps~\cite{HUANG2016,PRITYCHENKO2016}. An additional gate on the $5^-_1\rightarrow4^+_1$ transition detected in a HPGe detector was imposed that increased the quality of the LaBr$_3$(Ce) timing spectrum. Due to the improved peak-to-background ratio (now in the 20:1 range) and the abundant statistics, the centroids were measured with a precision of 2~ps, see Fig.~\ref{fig:Hg198_2+_lifetime}. 

\subsection{Deconvolution method}

When the measured lifetime is comparable or longer than the timing resolution of the system (FWHM$\sim 330$~ps for this experiment), the time distribution will present an exponential decay on the delayed part. The lifetime can be extracted directly from the slope of the decay delayed part. This time distribution can be fitted to a Gaussian convoluted with an exponential decay of the form:

\begin{equation}
  F(t_j) =\gamma \int_A^{+\infty} e^{- \delta (t_j - t)^2} e^{-\lambda t}dt \label{eq:convolution_method}
\end{equation}

\noindent where $\gamma$ is the normalization factor, $\delta$ is a parameter related to the width of the Gaussian prompt distribution and $A$ is the centroid of said Gaussian, which is related to the position of a prompt transition of the same energy. When needed, additional terms to account for the time-random background can be introduced.  Additional details on the method are given in Ref.~\cite{GAR19}.

The ZDS detector used as a TAC start signal is particularly well suited for this convolution method, by giving the time difference between the $\beta^-/\beta^+$ particle and the $\gamma$-ray. Thanks to its reduced timing FWHM, lifetimes will show a slope at shorter lifetime values. Moreover, since it detects charged particles, not $\gamma$-rays, it does not require a transition feeding the excited state of interest, it can be started by the $\beta$ particle directly populating the level. This allows access to levels that are unavailable to LaBr$_3$(Ce)-LaBr$_3$(Ce) coincidences. Lastly, because of its larger efficiency and the fact that it does not require a $\gamma$-ray cascade to operate, in general it will yield far superior statistics. In a similar fashion to the GCDM, additional HPGe coincidences can be imposed to increase selectivity. 

\section{Experimental results\label{sec:experimental_results}}

\begin{figure}
\begin{center}
\includegraphics[width=\columnwidth, keepaspectratio]{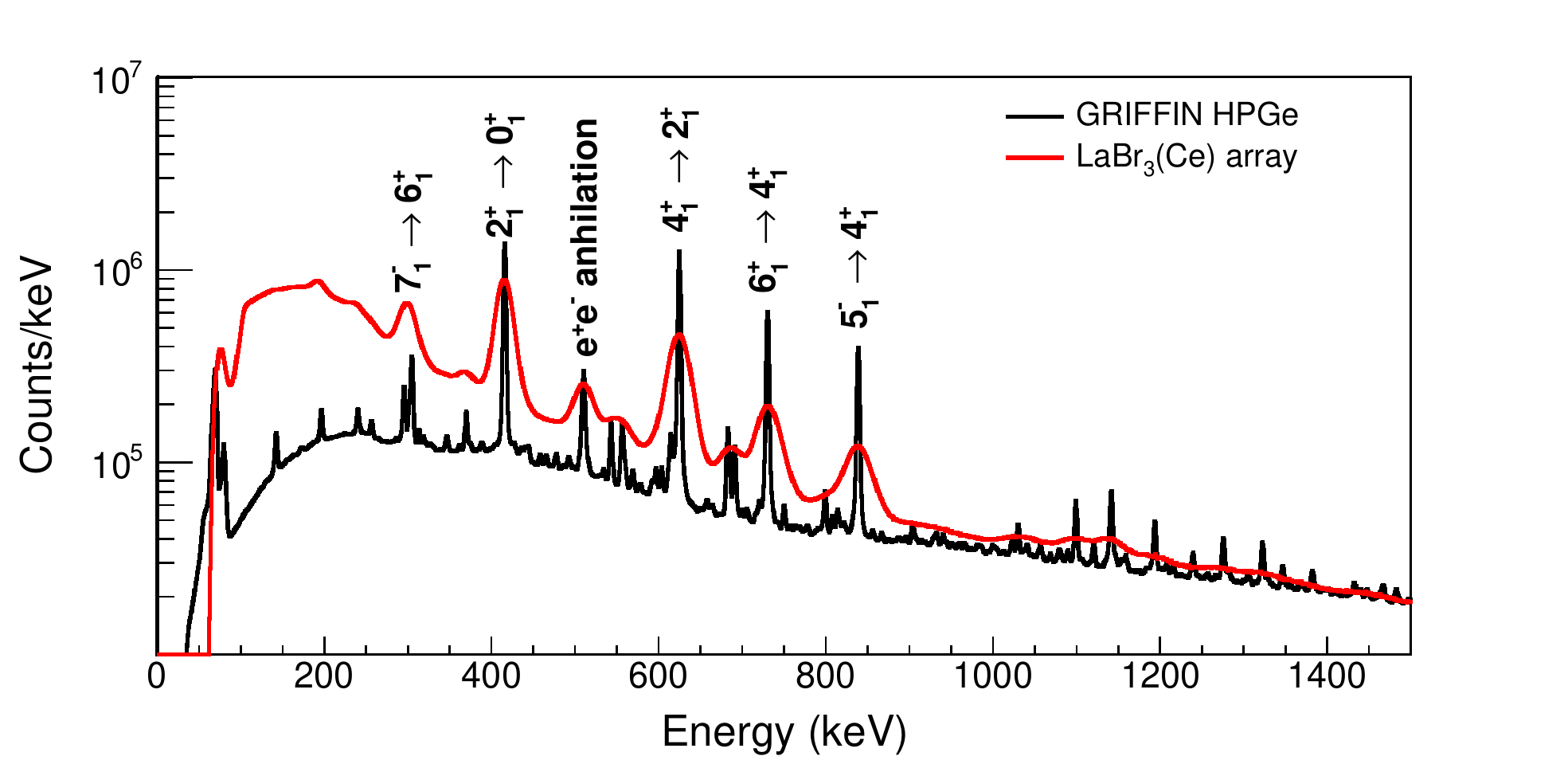}
\caption{\label{fig:energy_spectra} Energy spectra of the LaBr$_3$(Ce) (\textit{red/gray}) and HPGe GRIFFIN (\textit{black}) arrays for the decay of $^{190m}$Tl. This data was taken in LaBr$_3$(Ce)-LaBr$_3$(Ce)-TAC-(HPGe) coincidences mode. Some of the most intense transitions in $^{190}$Hg have been labeled.}
\end{center}
\end{figure}

In the present high-statistics study (see Fig.~\ref{fig:energy_spectra} for an example of the quality and quantity of data collected), a large number of lifetimes have been observed across the \textit{n}-deficient Hg isotopic chain. Table~\ref{tab:lifetimes} summarizes all the measured half-lives from this work using the GCDM described in Sec.~\ref{sec:data_analysis} and Ref.~\cite{Regis2013}. E$_\textrm{feeder}$ and E$_\textrm{decay}$ indicate the energies used for the gating transitions in the LaBr$_3$(Ce) crystals. E$_\textrm{HPGe}$ indicates the gated transition in the HPGe array. When previously measured, the literature value is given in the table for comparison. In some cases more than one combination of transition gating could measure the lifetime. In those cases, all employed combinations are described in the table, and as final result the average is given.

\begin{table*}
\caption{Summary of the half-lives measured in this experiment with comparison to values in literature. The level and transition energies are taken from Ref.~\cite{NNDC}. E$_\textrm{feeder}$ gives the energy of the feeding transition selected in the LaBr$_3$(Ce) crystals and E$_\textrm{decay}$ to the decaying one. E$_\textrm{HPGe}$ makes reference to the additional condition set in the HPGe detectors. When more than one energy is given in a column, it indicates that different combinations of transitions were used to obtain the lifetime of the level. No significant discrepancies were found between different combinations and the average value is given as the final result.}
\label{tab:lifetimes}
\begin{center}
	\begin{tabular}{cccccccccccc}
	\hline
Isotope	&	E$_\textrm{level}$	&	 $J^\pi$	&	E$_\textrm{feeder}$	&	 $J^\pi_{fi} \rightarrow \text{J}^\pi_{ff}$	&	E$_\textrm{decay}$ 	&	 $J^\pi_{di} \rightarrow \text{J}^\pi_{df}$	&	E$_\textrm{HPGe}$	&	$J^\pi_{gi} \rightarrow \text{J}^\pi_{gf}$	&	T$_{1/2}$ exp. 	&	T$_{1/2}$ lit. 	&	Ref	\\
	&	 (keV)	&		&	(keV)	&		&	(keV)	&		&	 (keV)	&		&	 (ps)	&	 (ps)	&		\\ \hline
$^{200}$Hg	&	367.9	&	$2^+_1$	&	579.3	&	$4^+_1 \rightarrow 2^+_1$	&	367.9	&	$2^+_1 \rightarrow 0^+_1$	&	828.3	&	$3^+_3 \rightarrow 4^+_1$	&	44(3)	&	46.4(4)	&	\cite{PRITYCHENKO2016}	\\
	&	947.2	&	$4^+_1$	&	828.3	&	$3^+_3 \rightarrow 4^+_1$	&	579.3	&	$4^+_1 \rightarrow 2^+_1$	&	367.9	&	$2^+_1 \rightarrow 0^+_1$	&	6(3)	&	3.24(5)	&	\cite{Bockisch1979, Gunther1981}	\\
	&	1029.3	&	$0^+_2$	&	701.6	&	$2^+_5 \rightarrow 0^+_2$	&	661.4	&	$0^+_2 \rightarrow 2^+_1$	&	367.9	&	$2^+_1 \rightarrow 0^+_1$	&	8(4)	&		&		\\
	&	1254.1	&	$2^+_2$	&	628.8	&	$2^+_7 \rightarrow 2^+_2$	&	886.2	&	$2^+_2 \rightarrow 2^+_1$	&	367.9	&	$2^+_1 \rightarrow 0^+_1$	&	8(6)	&	3.5(8)	&	\cite{Bockisch1979}	\\ \hline
$^{198}$Hg	&	411.8	&	$2^+_1$	&	636.7	&	$4^+_1 \rightarrow 2^+_1$	&	411.8	&	$2^+_1 \rightarrow 0^+_1$	&	587.2	&	$5^-_1 \rightarrow 4^+_1$	&	24(3)	&	23.15(28)	&	\cite{PRITYCHENKO2016}	\\
	&	1048.5	&	$4^+_1$	&	587.2	&	$5^-_1 \rightarrow 4^+_1$	&	636.7	&	$4^+_1 \rightarrow 2^+_1$	&	411.8	&	$2^+_1 \rightarrow 0^+_1$	&	<5	&	1.80(8)	&	\cite{Bockisch1979, Gunther1981}	\\
	&	1635.7	&	$5^-_1$	&	489.6	&	$(6,7)^-_1 \rightarrow 5^-_1$	&	587.2	&	$5^-_1 \rightarrow 4^+_1$	&	411.8	&	$2^+_1 \rightarrow 0^+_1$	&	57(7)	&	62(11)	&	\cite{Beraud1971}	\\
	&	1683.4	&	$7^-_1$	&	226.2	&	$6^-_1 \rightarrow 7^-_1$	&	587.2	&	$5^-_1 \rightarrow 4^+_1$	&	411.8	&	$2^+_1 \rightarrow 0^+_1$	&	6.6(1) ns	&	6.9(2) ns	&	\cite{DULFER1970}	\\ \hline
$^{196}$Hg	&	426.0	&	$2^+_1$	&	635.5	&	$4^+_1 \rightarrow 2^+_1$	&	426.0	&	$2^+_1 \rightarrow 0^+_1$	&	695.6	&	$5^-_1 \rightarrow 4^+_1$	&	16(3)	&	17.2(6)	&	\cite{PRITYCHENKO2016}	\\
	&	1061.4	&	$4^+_1$	&	695.6	&	$5^-_1 \rightarrow 4^+_1$	&	635.5	&	$4^+_1 \rightarrow 2^+_1$	&	426.0	&	$2^+_1 \rightarrow 0^+_1$	&	4(3)	&	4(3)	&	\cite{Esmaylzadeh2018}	\\
	&	1757.0	&	$5^-_1$	&	588.8	&	$(5,6,7)^-_1 \rightarrow 5^-_1$	&	695.6	&	$5^-_1 \rightarrow 4^+_1$	&	426.0	&	$2^+_1 \rightarrow 0^+_1$	&	670(80)	&	555(17)	&	\cite{TON1970}	\\
	&	1841.3	&	$7^-_1$	&	505.2	&	$(5,6,7)^-_1 \rightarrow 7^-_1$	&	695.6	&	$5^-_1 \rightarrow 4^+_1$	&	426.0	&	$2^+_1 \rightarrow 0^+_1$	&	4.8(2) ns	&	5.22(16) ns	&	\cite{TON1970}	\\ \hline
$^{194}$Hg	&	427.9	&	$2^+_1$	&	636.3	&	$4^+_1 \rightarrow 2^+_1$	&	427.9	&	$2^+_1 \rightarrow 0^+_1$	&	734.8	&	$6^+_1 \rightarrow 4^+_1$	&	19(1)	&	15(3)	&	\cite{Esmaylzadeh2018}	\\
	&		&		&		&		&		&		&	748.9	&	$5^-_1 \rightarrow 4^+_1$	&		&		&		\\
	&	1064.2	&	$4^+_1$	&	734.8	&	$6^+_1 \rightarrow 4^+_1$	&	636.3	&	$4^+_1 \rightarrow 2^+_1$	&	427.9	&	$2^+_1 \rightarrow 0^+_1$	&	<3	&	5(3)	&	\cite{Esmaylzadeh2018}	\\
	&	1813	&	$5^-_1$	&	650.3	&	$6^-_2 \rightarrow 5^-_1$	&	748.9	&	$5^-_1 \rightarrow 4^+_1$	&	427.9	&	$2^+_1 \rightarrow 0^+_1$	&	51(6)	&	<150	&	\cite{TON1970}	\\
	&	1910.0	&	$7^-_1$	&	255.4	&	$6^-_1 \rightarrow 7^-_1$	&	734.8	&	$6^+_1 \rightarrow 4^+_1$	&	427.9	&	$2^+_1 \rightarrow 0^+_1$	&	3.46(3) ns	&	3.75(11) ns	&	\cite{Gunther1977}	\\
	&		&		&	208.9	&	$(6,7,8)^-_1 \rightarrow 6^-_1$	&	748.9	&	$5^-_1 \rightarrow 4^+_1$	&		&		&		&		&		\\
	&		&		&		&		&	111.0	&	$7^-_1 \rightarrow 6^+_1$	&		&		&		&		&		\\ \hline
$^{192}$Hg	&	422.8	&	$2^+_1$	&	634.8	&	$4^+_1 \rightarrow 2^+_1$	&	422.8	&	$2^+_1 \rightarrow 0^+_1$	&	786.0	&	$5^-_1 \rightarrow 4^+_1$	&	12(1)	&	15(6)	&	\cite{Esmaylzadeh2018}	\\
	&		&		&		&		&		&		&	745.5	&	$6^+_1 \rightarrow 4^+_1$	&		&		&		\\
	&	1057.6	&	$4^+_1$	&	745.5	&	$6^+_1 \rightarrow 4^+_1$	&	634.8	&	$4^+_1 \rightarrow 2^+_1$	&	422.8	&	$2^+_1 \rightarrow 0^+_1$	&	4(3)	&	4(3)	&	\cite{Esmaylzadeh2018}	\\
	&	1803.0	&	$6^+_1$	&	174.0	&	$7^-_1 \rightarrow 6^+_1$	&	745.5	&	$6^+_1 \rightarrow 4^+_1$	&	634.8	&	$4^+_1 \rightarrow 2^+_1$	&	73(10)	&		&		\\
	&	1843.9	&	$5^-_1$	&	133.1	&	$7^-_1 \rightarrow 5^-_1$	&	786.3	&	$5^-_1 \rightarrow 4^+_1$	&	634.8	&	$4^+_1 \rightarrow 2^+_1$	&	383(14)	&		&		\\
	&	1977.0	&	$7^-_1$	&	239.2	&	$8^-_1 \rightarrow 7^-_1$	&	174.0	&	$7^-_1 \rightarrow 6^+_1$	&	422.8	&	$2^+_1 \rightarrow 0^+_1$	&	1.03(5) ns	&	1.04(6) ns	&	\cite{MERTIN1978}	\\
	&		&		&		&		&		&		&	745.5	&	$6^+_1 \rightarrow 4^+_1$	&		&		&		\\ \hline
$^{190}$Hg	&	416.3	&	$2^+_1$	&	625.4	&	$4^+_1 \rightarrow 2^+_1$	&	416.4	&	$2^+_1 \rightarrow 0^+_1$	&	731.1	&	$6^+_1 \rightarrow 4^+_1$	&	15(1)	&	15(6)	&	\cite{Esmaylzadeh2018}	\\
	&		&		&		&		&		&		&	839.7	&	$5^-_1 \rightarrow 4^+_1$	&		&		&		\\
	&	1041.8	&	$4^+_1$	&	731.1	&	$6^+_1 \rightarrow 4^+_1$	&	625.4	&	$4^+_1 \rightarrow 2^+_1$	&	416.4	&	$2^+_1 \rightarrow 0^+_1$	&	5(4)	&	<8	&	\cite{Esmaylzadeh2018}	\\
	&		&		&	839.6	&	$5^-_1 \rightarrow 4^+_1$	&		&		&		&		&		&		&		\\
	&	1772.9	&	$6^+_1$	&	305.4	&	$7^-_1 \rightarrow 6^+_1$	&	416.4	&	$2^+_1 \rightarrow 0^+_1$	&	625.4	&	$4^+_1 \rightarrow 2^+_1$	&	7(4)	&		&		\\
	&	1881.2	&	$5^-_1$	&	370.3	&	$(6,7)^-_1 \rightarrow 5^-_1$	&	839.6	&	$5^-_1 \rightarrow 4^+_1$	&	625.4	&	$4^+_1 \rightarrow 2^+_1$	&	<40	&		&		\\
	&		&		&	196.9	&	$7^-_1 \rightarrow  5^-_1$	&		&		&		&		&		&		&		\\
	&	2078.3	&	$7^-_1$	&	240.6	&	$8^-_1 \rightarrow 7^-_1$	&	305.4	&	$7^-_1 \rightarrow 6^+_1$	&	731.0	&	$6^+_1 \rightarrow 4^+_1$	&	<200	&		&		\\ \hline
$^{188}$Hg	&	412.9	&	$2^+_1$	&	592.1	&	$4^+_1 \rightarrow 2^+_1$	&	412.9	&	$2^+_1 \rightarrow 0^+_1$	&	504.3	&	$6^+_1 \rightarrow 4^+_1$	&	14(3)	&	13.1(21)	&	\cite{Bree2014}	\\
	&		&		&		&		&		&		&	772.4	&	$6^+_2 \rightarrow 4^+_1$	&		&		&		\\
	&		&		&		&		&		&		&	904.8	&	$5^-_1 \rightarrow 4^+_1$	&		&		&		\\
	&	881.1	&	$2^+_2$	&	326.9	&	$4^+_2 \rightarrow 2^+_2$	&	468.2	&	$2^+_2 \rightarrow 2^+_1$	&	412.9	&	$2^+_1 \rightarrow 0^+_1$	&	<20	&	141(31)	&	\cite{JOSHI1994}	\\
	&	1004.9	&	$4^+_1$	&	504.3	&	$6^+_1 \rightarrow 4^+_1$	&	592.1	&	$4^+_1 \rightarrow 2^+_1$	&	412.9	&	$2^+_1 \rightarrow 0^+_1$	&	<30	&	1.60(12)	&	\cite{Bree2014}	\\
	&	1207.9	&	$4^+_2$	&	700.1	&	$(4,5)^+ \rightarrow 4^+_2$	&	326.9	&	$4^+_2 \rightarrow 2^+_2$	&	412.9	&	$2^+_1 \rightarrow 0^+_1$	&	<40	&		&		\\
	&		&		&		&		&	795.2	&	$4^+_2 \rightarrow 2^+_1$	&		&		&		&		&		\\
	&	1509.2	&	$6^+_1$	&	460.7	&	$8^+_1 \rightarrow 6^+_1$	&	504.3	&	$6^+_1 \rightarrow 4^+_1$	&	592.1	&	$4^+_1 \rightarrow 2^+_1$	&	<10	&		&		\\
	&	1777.2	&	$6^+_2$	&	424.1	&	$7^-_1 \rightarrow 6^+_1$	&	772.4	&	$6^+_2 \rightarrow 4^+_1$	&	412.9	&	$2^+_1 \rightarrow 0^+_1$	&	<30	&		&		\\
	&	1909.7	&	$5^-_1$	&	385.8	&	$6^-_1 \rightarrow 5^-_1$	&	904.8	&	$5^-_1 \rightarrow 4^+_1$	&	592.1	&	$4^+_1 \rightarrow 2^+_1$	&	10(9)	&		&		\\ \hline

\end{tabular}
\end{center}
\end{table*}

With the exception of the $7^-_1$ state in $^{190}$Hg (see below), lifetimes given as upper limits are the results of uncertainties larger than the measured values. The main reason for this is a large peak-to-background ratio for weak transitions decaying from a level with a short half-life. When this ratio is $\lesssim 1$, the Compton contribution is substantial and the uncertainty induced by the correction from Eq.~\ref{eq:compton_corr} is generally larger than the lifetime.

\begin{figure}
\begin{center}
\includegraphics[width=\columnwidth, keepaspectratio]{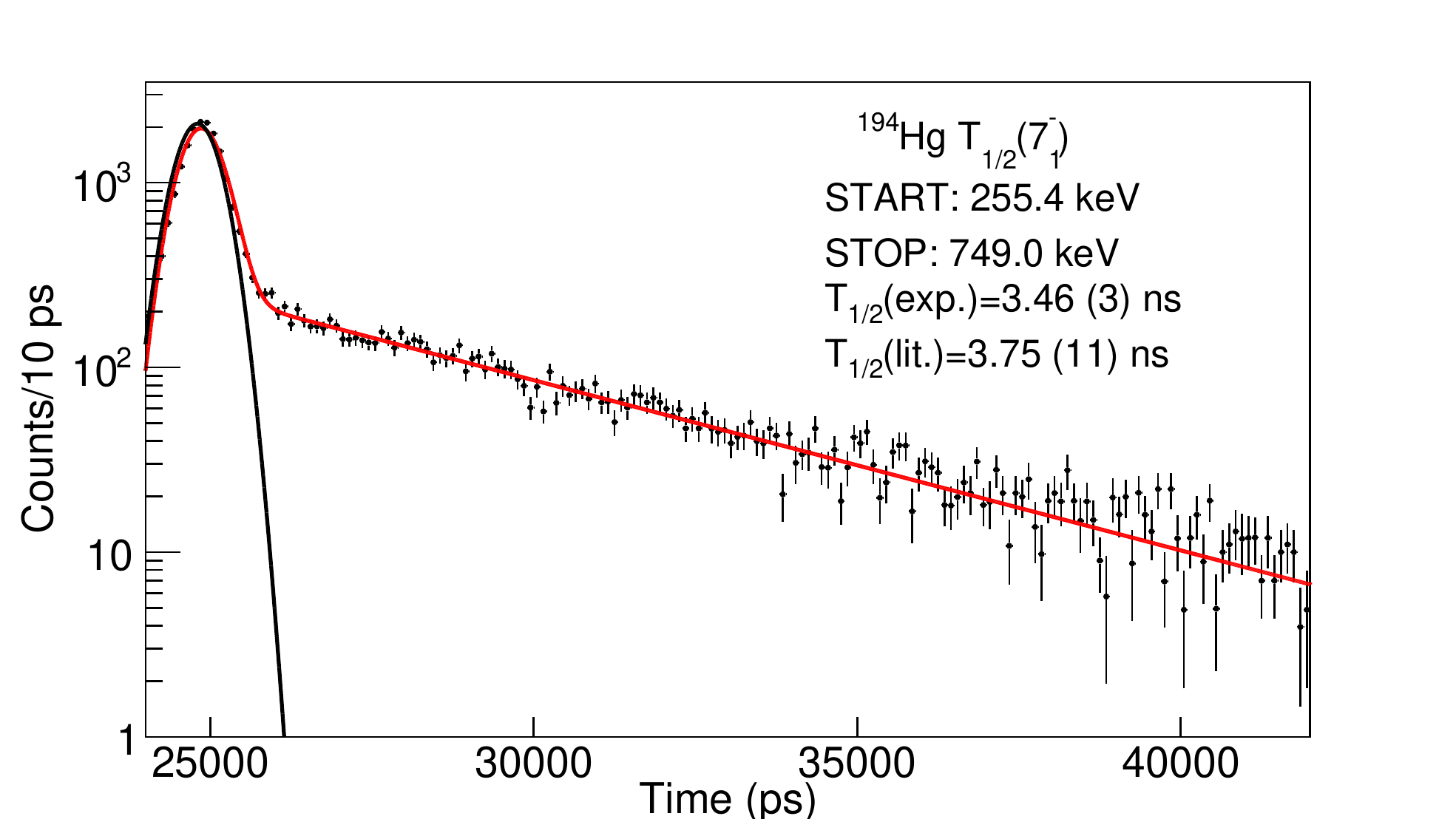}
\caption{\label{fig:Hg194_7-_lifetime} Lifetime of the $7^-_1$ state in $^{194}$Hg. The fit was performed using a Gaussian convoluted with a double exponential decay, to account for the short lifetime background under the peak, plus a constant term for the time-random background.}
\end{center}
\end{figure}

For the purpose of this work, lifetimes longer than 100~ps have been obtained by the convolution method, fitting Eq.~\ref{eq:convolution_method} to the time distribution. Every lifetime was measured in delayed (positive lifetime) and anti-delayed (negative lifetime) coincidences. Since these are physically different events, every half-life was effectively measured twice. The result shown in Tab.~\ref{tab:lifetimes} is the average of the two values, which in all cases were in good agreement. Figure~\ref{fig:Hg194_7-_lifetime} shows an example of a lifetime extracted using this method. The $6^-_1 \rightarrow 7^-_1$ transition in $^{194}$Hg is used as START and the $5^-_1 \rightarrow 4^+_1$ as STOP. The $7^-_1$ state decays to the $5^-_1$ state with a 56\% branching ratio and the $5^-_1$ state lifetime is known to be in the picoseconds range (see Table~\ref{tab:lifetimes}), so the long tail shown in Fig.~\ref{fig:Hg194_7-_lifetime} can be unambiguously attributed to the $7^-_1$ state. The time distribution was fitted to a Gaussian function convoluted to a double exponential decay. This second exponential decay is introduced to account for the significant background, which in this case has a much shorter lifetime. An additional constant term was introduced to fit the time-random background.

The lifetime of the $7^-_1$ state in $^{190}$Hg was estimated using a different method. The 305.4-keV $7^-_1 \rightarrow 6^+_1$ decaying transition was visible and could be selected in the LaBr$_3$(Ce) spectrum. An additional transition, 731.1-keV $6^+_1 \rightarrow 4^+_1$, was selected in the HPGe array to reduce the background. Since no discernible feeding transition could be used as the gating transition in the LaBr$_3$(Ce) detectors under these conditions, the lifetime was extracted from ZDS-LaBr$_3$(Ce)-HPGe coincidences. This time difference between the $\beta^+$ particle and the $7^-_1 \rightarrow 6^+_1$ transition was a composition of the lifetimes of the $7^-_1$ and all the levels feeding it from above, which in this case are the $(8^-)$ and $(9^-)$ states~\cite{NNDC}. The resulting TAC spectrum showed no delayed component and thus a conservative upper limit of T$_{1/2}<200$~ps was estimated from the FWHM  and the lack of slope. From previous experiments (see Refs.~\cite{Bingham1976, Kortelahti1991, DEL94, NNDC}) it has been established that the decay of $^{190m}$Tl favors the $7^-$ state over the $(8^-)$ and $(9^-)$ levels in $\sim 75\%$ of the decays. For this reason, it cannot be discarded that the $(8^-)$ or $(9^-)$ states have a long lifetime which could not be observed under these conditions, therefore no limits have been deduced for them.

\begin{figure}
\begin{center}
\includegraphics[width=\columnwidth, keepaspectratio]{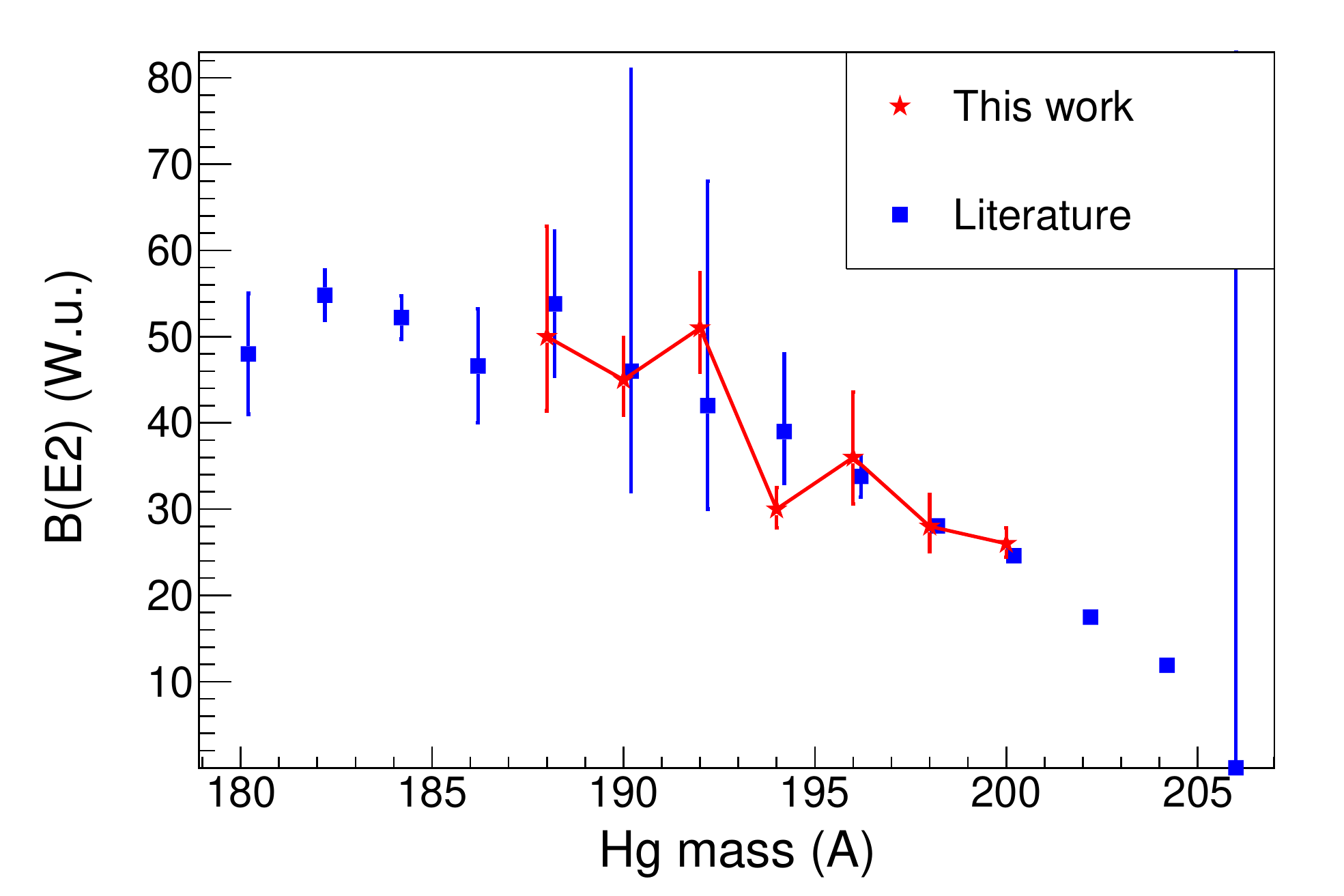}
\caption{\label{fig:be2_2} Comparison of the experimentally deduced B(E2;$2^+_1 \rightarrow 0^+_1$) values from this work and previous literature values. Literature values are taken from the evaluated compilation~\cite{PRITYCHENKO2016}, with the exception of $^{190,192,194}$Hg, which are taken from Ref.~\cite{Esmaylzadeh2018}.  Note that the error for the $^{200}$Hg B(E2) value evaluated in~\cite{PRITYCHENKO2016} has one too many digits, making it 10 times larger than it should be. This has been corrected in the present plot.}
\end{center}
\end{figure}

With the exception of the lifetime of $2^+_2$ in $^{188}$Hg (discussed in Sec.~\ref{sec:188Hg-2_2-lifetime}) there is excellent agreement between the results obtained in this work and previous measurements (see Fig.~\ref{fig:be2_2} and Table~\ref{tab:lifetimes}). In particular, excellent agreement is observed with the lifetimes of the $2_1^+$ states of $^{196,198,200}$Hg stable isotopes for which lifetimes have been measured a number of times and are precisely known. This is a strong validation of the quality of the results and the ability of the LaBr$_3$(Ce)-GRIFFIN array to measure lifetimes in the picoseconds range using the GCDM. It should be noted here that the $^{198}$Hg $2_1^+$ state literature lifetime was used to calibrate the time walk of the array and reduce the uncertainty of all the other measured lifetimes, but was not used when determining its own value in this work. The agreement for the $7^-_1$ state lifetimes,  measured using the convolution method, is poorer than with the other states. For $^{194,196,198}$Hg these values are $\sim 2\sigma$ lower than previous measurements. No satisfactory explanation for this  deviation was found.

\begin{table*}
\caption{Summary of the deduced reduced probabilities from the results of this experiment. For $\Delta$J=0, $\Delta\pi=0$ transitions, the deduced B(M1) and B(E2) values are given assuming pure multipolarities. The exception is the 886.9-keV $2^+_2 \rightarrow 2^+_1$ transition in $^{200}$Hg for which a $\delta^2=-2.20(10)$ was measured previously and has been used in the calculations~\cite{Breitig1974, Ahmad1989}. Energies and branching ratios are taken from Ref.~\cite{NNDC}. All values have been corrected by the conversion electron coefficient from Ref.~\cite{BRICC}.}
\label{tab:reduced_probabilities}
\begin{center}
	\begin{tabular}{ccccccccc}
	\hline
Isotope	&	J$^\pi_i$	&	T$_{1/2}$	&	J$^\pi_f$	&	E$_\gamma$	&	$\rho^2$(E0)$\times 10^3$	&	B(E1)	&	B(M1)	&	B(E2)	\\ 	
	&		&	(ps)	&		&	(keV)	&		&	W.u.	&	W.u.	&	W.u.	\\ \hline	
$^{200}$Hg	&	$2^+_1$	&	44(3)	&	$0^+_1$	&	367.9	&		&		&		&	26(2)	\\	
	&	$4^+_1$	&	6(3)	&	$2^+_1$	&	579.3	&		&		&		&	20(9)	\\	
	&	$0^+_2$	&	8(4)	&	$2^+_1$	&	661.4	&		&		&		&	8(4)	\\	
	&		&		&	$0^+_1$	&	1029.3	&	0.02(1)	&		&		&		\\	
	&	$2^+_2$	&	8(6)	&	$0^+_2$	&	224.8	&		&		&		&	4(3)	\\	
	&		&		&	$4^+_1$	&	306.9	&		&		&		&	0.6(5)	\\	
	&		&		&	$2^+_1$	&	886.2	&		&		&	$5(4) \times 10^{-4}$	&	1.0(8)	\\	
	&		&		&	$0^+_1$	&	1254.1	&		&		&		&	0.10(8)	\\ \hline	
$^{198}$Hg	&	$2^+_1$	&	24(3)	&	$0^+_1$	&	411.8	&		&		&		&	28(4)	\\	
	&	$4^+_1$	&	<5	&	$2^+_1$	&	636.7	&		&		&		&	>16	\\	
	&	$5^-_1$	&	57(7)	&	$4^+_1$	&	587.2	&		&	$1.7(2) \times 10^{-5}$	&		&		\\	
	&	$7^-_1$	&	6.6(1) ns	&	$5^-_1$	&	47.7	&		&		&		&	29.5(5)	\\ \hline	
$^{196}$Hg	&	$2^+_1$	&	16(3)	&	$0^+_1$	&	426.0	&		&		&		&	36(7)	\\	
	&	$4^+_1$	&	4(3)	&	$2^+_1$	&	635.5	&		&		&		&	20(15)	\\	
	&	$5^-_1$	&	670(80)	&	$4^+_1$	&	695.6	&		&	$8.9(11) \times 10^{-7}$	&		&		\\	
	&	$7^-_1$	&	4.8(2) ns	&	$5^-_1$	&	84.3	&		&		&		&	33(1)	\\ \hline	
$^{194}$Hg	&	$2^+_1$	&	19(1)	&	$0^+_1$	&	427.9	&		&		&		&	30(2)	\\	
	&	$4^+_1$	&	<3	&	$2^+_1$	&	636.3	&		&		&		&	>27	\\	
	&	$5^-_1$	&	51(6)	&	$4^+_1$	&	748.9	&		&	$4.0(1) \times 10^{-6}$	&		&		\\	
	&	$7^-_1$	&	3.46(3) ns	&	$5^-_1$	&	97.0	&		&		&		&	34.4(3)	\\	
	&		&		&	$6^+_1$	&	111.0	&		&	$1.4(5) \times 10^{-5}$	&		&		\\ \hline	
$^{192}$Hg	&	$2^+_1$	&	12(1)	&	$0^+_1$	&	422.8	&		&		&		&	51(4)	\\	
	&	$4^+_1$	&	4(3)	&	$2^+_1$	&	634.8	&		&		&		&	21(15)	\\	
	&	$6^+_1$	&	73(10)	&	$4^+_1$	&	745.5	&		&		&		&	0.51(7)	\\	
	&	$5^-_1$	&	383(14)	&	$4^+_1$	&	786.3	&		&	$1.7(1) \times 10^{-5}$	&		&		\\	
	&	$7^-_1$	&	1.03(5) ns	&	$5^-_1$	&	133.1	&		&		&		&	37(2)	\\	
	&		&		&	$6^+_1$	&	174.0	&		&	$2.38(3) \times 10^{-5}$	&		&		\\ \hline	
$^{190}$Hg	&	$2^+_1$	&	15(1)	&	$0^+_1$	&	416.4	&		&		&		&	45(3)	\\	
	&	$4^+_1$	&	5(4)	&	$2^+_1$	&	625.4	&		&		&		&	18(14)	\\	
	&	$6^+_1$	&	7(4)	&	$4^+_1$	&	731.1	&		&		&		&	6(3)	\\	
	&	$5^-_1$	&	<40	&	$4^+_1$	&	839.6	&		&	$>6 \times 10^{-6}$	&		&		\\	
	&	$7^-_1$	&	<200	&	$5^-_1$	&	196.9	&		&		&		&	>30	\\	
	&		&		&	$6^+_1$	&	305.4	&		&	$>2.4 \times 10^{-5}$	&		&		\\ \hline	
$^{188}$Hg	&	$2^+_1$	&	14(3)	&	$0^+_1$	&	412.9	&		&		&		&	50(11)	\\	
	&	$2^+_2$	&	<20	&	$2^+_1$	&	468.2	&		&		&	$>4 \times 10^{-3}$	&	>8	\\	
	&		&		&	$0^+_1$	&	881.1	&		&		&		&	>1	\\	
	&	$4^+_1$	&	<30	&	$2^+_1$	&	592.1	&		&		&		&	>4	\\	
	&	$4^+_2$	&	<40	&	$4^+_1$	&	203.2	&		&		&	$>3 \times 10^{-3}$	&	>33	\\	
	&		&		&	$2^+_2$	&	326.9	&		&		&		&	>26	\\	
	&		&		&	$2^+_1$	&	795.2	&		&		&		&	>0.32	\\	
	&	$6^+_1$	&	<10	&	$(4^+_4)$	&	269.4	&		&		&		&	>110	\\	
	&		&		&	$4^+_2$	&	301.2	&		&		&		&	>250	\\	
	&		&		&	$4^+_1$	&	504.3	&		&		&		&	>90	\\	
	&	$6^+_2$	&	<30	&	$4^+_2$	&	569.3	&		&		&		&	>1.1	\\	
	&		&		&	$4^+_1$	&	772.4	&		&		&		&	>0.8	\\	
	&	$5^-_1$	&	10(9)	&	$4^+_2$	&	701.7	&		&	$4.1(37) \times 10^{-6}$	&		&		\\ 	
	&		&		&	$4^+_1$	&	904.8	&		&	$2.6(23) \times 10^{-5}$	&		&		\\ \hline	

\end{tabular}
\end{center}
\end{table*}

\section{Calculations and discussion}

\subsection{Positive-parity yrast states}

The \textit{Z}$\sim$82, \textit{N}$\sim 104$ mid-shell nuclei are beyond reach of most shell-model calculations, with only the state-of-the-art Monte Carlo shell model (MCSM)~\cite{Otsuka2001} having been recently used to calculate some of the most basic properties of the ground and first excited states of $^{177-186}$Hg~\cite{Marsh2018, Sels2019}. In order to study more complex properties of excited states, such as the B(E2) transition strengths, calculations must be carried out in truncated spaces. In the present study, we turn to results of the Interacting Boson Model (IBM) calculations. Two main sets of theoretical results are available for the B(E2) values of the neutron-deficient Hg isotopes; from the IBM-2 calculations~\cite{Nomura2013} and  IBM calculations that incorporate configuration mixing (IBM-CM) ~\cite{Garcia-Ramos2014}.

\begin{figure}
\begin{center}
\includegraphics[width=\columnwidth, keepaspectratio]{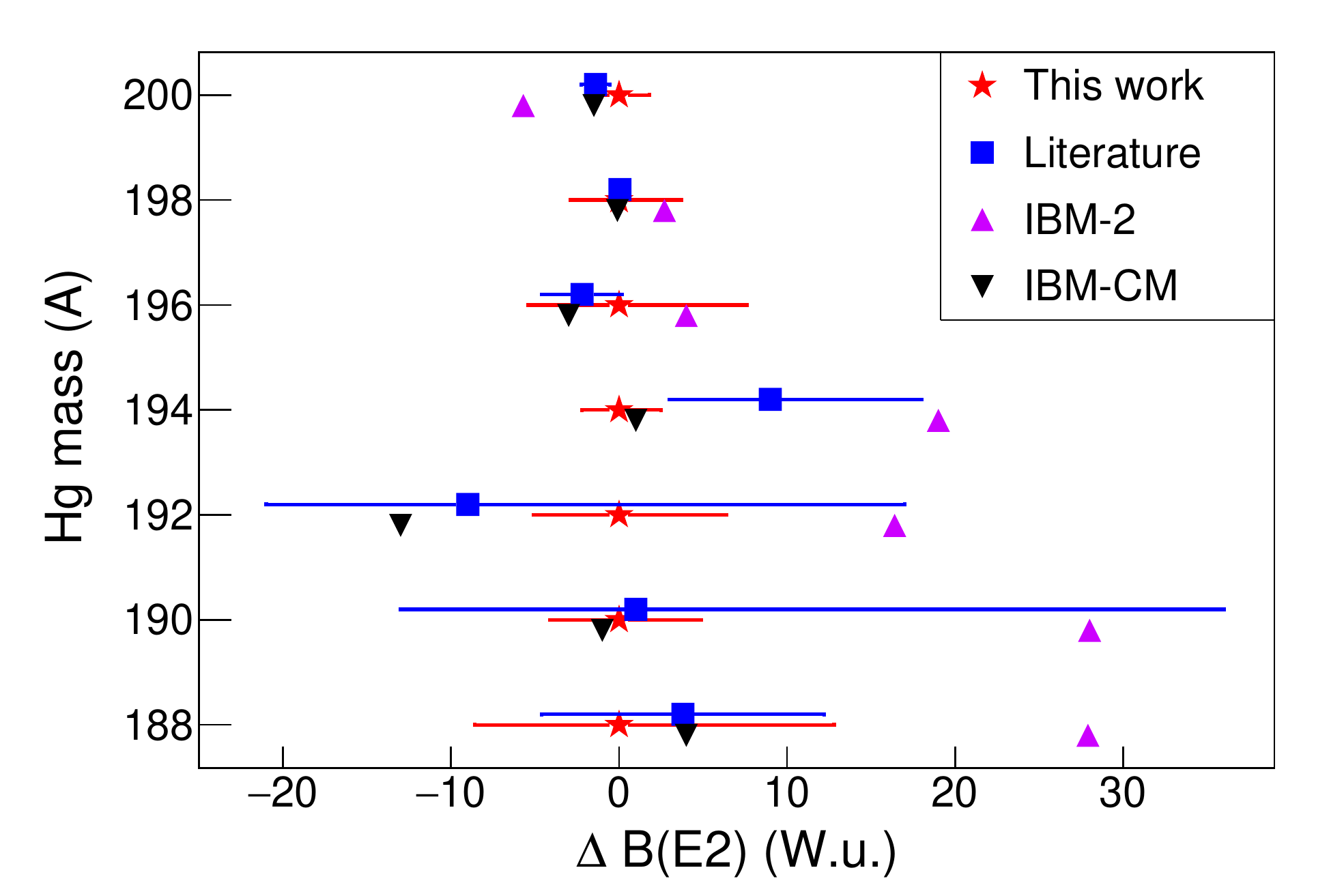}
\caption{\label{fig:exp_vs_theo_2state} Difference between the B(E2;$2^+_1 \rightarrow 0^+_1$) values measured in this work, literature and different IBM calculations. Literature values are taken from the evaluated compilation~\cite{PRITYCHENKO2016}, with the exception of $^{190,192,194}$Hg, which are taken from Ref.~\cite{Esmaylzadeh2018}. IBM-2 are from Ref.~\cite{Nomura2013} and IBM-CM from~\cite{Garcia-Ramos2014}.}
\end{center}
\end{figure}

\begin{figure}
\begin{center}
\includegraphics[width=\columnwidth, keepaspectratio]{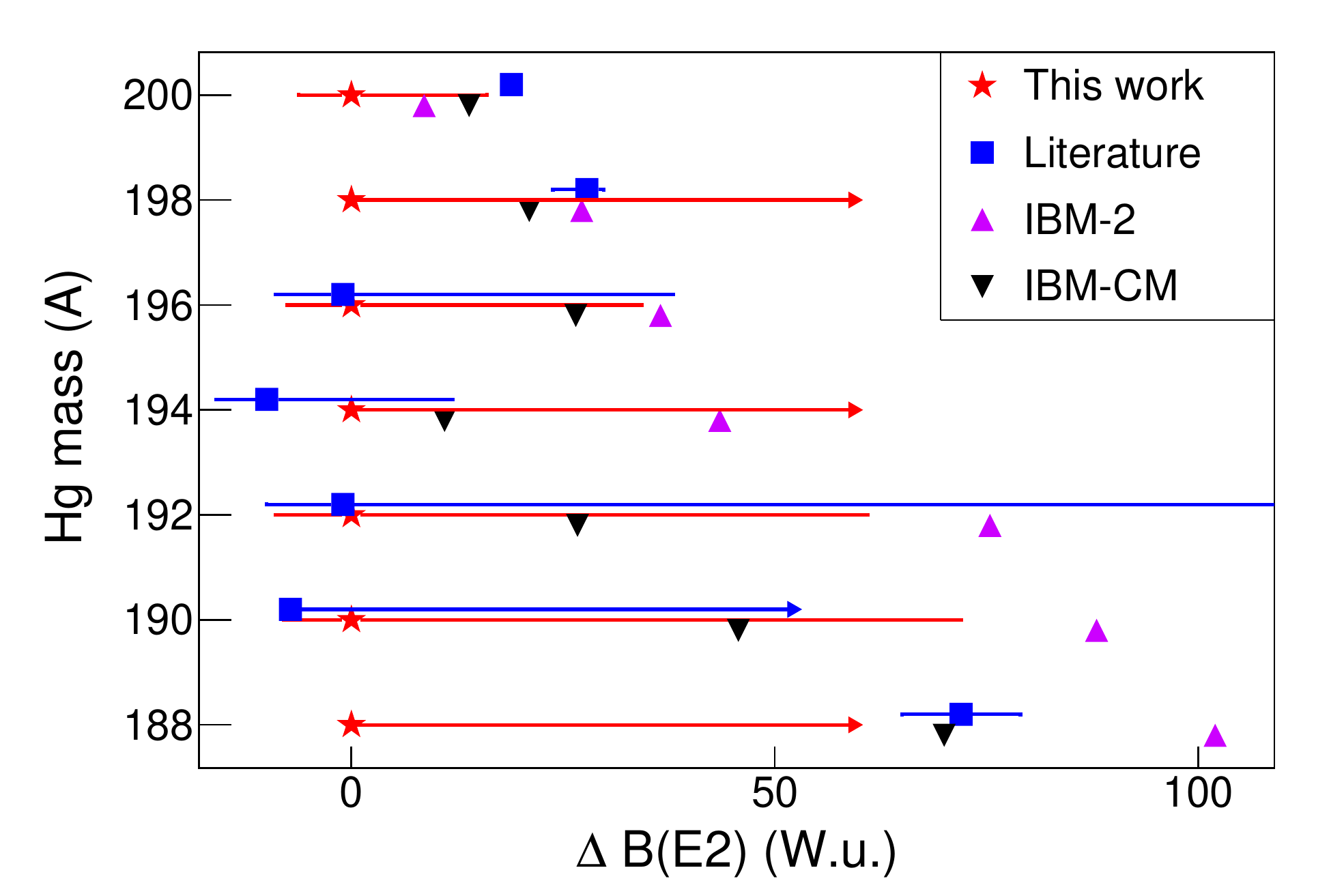}
\caption{\label{fig:exp_vs_theo_4state} Same as Fig.~\ref{fig:exp_vs_theo_2state}, but for the  B(E2;$4^+_1 \rightarrow 2^+_1$) values. }
\end{center}
\end{figure}

While both of them are based on the IBM, there are important differences between the two approaches. The IBM-2 calculations treat protons and neutrons independently, as opposed to the traditional IBM calculations (including IBM-CM), which makes no distinction between the different types of nucleon. Both calculations allow for proton excitations across \textit{Z}=82 and the formation of $4h-2p$ (intruder) states and mixing between the normal and intruder configurations. But while IBM-CM includes cross-shell excitations for the whole isotopic chain, the IBM-2 calculations that were performed in Ref.~\cite{Nomura2013} limits the inclusion of intruder states to the mass $^{172-190}$Hg isotopes, and uses a single configuration elsewhere. The other main difference between the two sets of calculations is that IBM-CM fitted the parameters to the (at the time) measured energies and B(E2) of the $2^+_1$ states, while the IBM-2 did not fit to any of the experimental results but mapped the IBM-2 Hamiltonian to results from a self-consistent mean field calculation using a Gogny-D1M energy-density functional.

Figures~\ref{fig:exp_vs_theo_2state} and~\ref{fig:exp_vs_theo_4state} show the difference between B(E2) values deduced from the lifetimes of the $2^+_1$ and $4^+_1$ states measured in the present work and theoretical calculations. Both the theoretical calculations and the experimental results show a smooth increase in B(E2) as neutrons are removed. In $^{200}$Hg, the B(E2;$2^+_1 \rightarrow 0^+_1$) value is $\sim 25$~W.u., which yields a moderate deformation of $\beta_2 = 0.098(2)$. This deformation increases as the mid-shell $N=104$ is approached, reaching a maximum around $^{182}$Hg with $\beta_2 = 0.147(4)$~\cite{PRITYCHENKO2016}. The new results are fully consistent with this picture but greatly improve the precision of these measurements for $^{190,192,194}$Hg.

Esmaylzadeh and collaborators~\cite{Esmaylzadeh2018} claimed a better agreement between the B(E2;$2^+_1 \rightarrow 0^+_1$) values they measured and the IBM-2 calculations, mainly because of the discrepancy they observed for $^{194}$Hg. The new, more precise values, indicate the opposite conclusion, a significantly better agreement with the IBM-CM is found, rather than with the IBM-2. The agreement between the IBM-CM predictions and the new data are well within one $\sigma$, with the exception of $^{192}$Hg. In this case, the B(E2;$2^+_1 \rightarrow 0^+_1$) value of $^{192}$Hg seems to have been underestimated by the IBM-CM and overestimated by the IBM-2 calculations. 

It is significant that the dip experimentally observed in the B(E2;$2^+_1 \rightarrow 0^+_1$) value of $^{194}$Hg is nicely reproduced when the configuration mixing is included in the calculations (IBM-CM clearly reproduces the staggering while IBM-2, which has no configuration mixing for this mass, does not). The IBM calculations do not include any type of sub-shell structure, but they are still able to reproduce this irregularity in the otherwise smooth evolution of the B(E2;$2^+_1 \rightarrow 0^+_1$) values.  While subtle effects can, of course, be introduced through the fitting of the IBM parameters to the energies of the states, it is nonetheless remarkable that the staggering of the B(E2) values is reproduced so well. (When the fits in Ref.~\cite{Garcia-Ramos2014} were done, none of the relevant B(E2) values in $^{190-194}$Hg were known.) A similar dip is observed in the evolution of the B(E2;$4^+_1 \rightarrow 2^+_1$) values also for $^{194}$Hg, where the lifetime measured by Esmaylzadeh and collaborators~\cite{Esmaylzadeh2018} certainly hints to the possibility of this staggering being present also for the $4^+_1$ state systematic.

On the other hand, the upper limits obtained for the lifetimes of the $4^+_1$ states (lower limits for the B(E2;$4^+_1 \rightarrow 2^+_1$)), are not stringent enough to distinguish between either of the models. The value for $^{192}$Hg (and maybe $^{196}$Hg) presented in this work and the value for $^{194}$Hg presented in~\cite{Esmaylzadeh2018} seem to favor the results from IBM-CM calculations, but no definitive conclusion can be achieved. More precise measurements of these lifetimes are required for a full description of the nuclei.

It is important to note that when the IBM-CM calculations were performed~\cite{Garcia-Ramos2014}, no lifetime information for $^{190,192,194}$Hg existed, so their B(E2) values were not included in the normalization or constraining of the calculations. It is, thus, remarkable, how the predictions of this set of calculations fit the measured values for $^{190}$Hg and $^{194}$Hg. The IBM-2 is not adjusted to experimental data as it is based on the fully-microscopic energy density functional calculation, so its ability to reproduce the general trend of the B(E2) values is significant. The great predictive power of these IBM calculations seems to validate their results of the minimal mixing between normal and intruder configurations for $^{192}$Hg and heavier isotopes. This confirms the hypothesis of studying these isotopes to benchmark the normal configuration free of perturbations from the intruder one, which in turn should facilitate the study of shape coexistance in the lighter ones.


\subsection{Negative-parity band \label{sec:negative_band}}

The yrast negative-parity band in the Hg isotopes has been successfully explained with a model of two quasi-particles coupled to an oblate rotor. One of the quasi-particles has a high spin ($11/2$ or $13/2$, from the $\pi h_{11/2}$ or the $\nu i_{13/2}$ orbitals, respectively) with a spherical wave function, while the other quasi-particle is of low spin ($\leq 5/2$, from the $pf$ shell) with a deformed Nilson wave function which is the combination of several configurations~\cite{Flaum1974,Neergard1975,Toki1977,Levon2006}.

In contrast, Beuscher \textit{et al.}~\cite{Beuscher1974} concluded that these structures are collective rotational bands involving many particles, not just the suggested two-particle coupling, because states up to $15^-$ were observed. This high spin can not be formed with just two-particles coupled to a core.

The lack of an intense $\gamma$ ray feeding the $7^-$ state in $^{190}$Hg prevented a precise measurement of the lifetime. Only a limit of $<200$~ps was obtained, which fits with the systematic of the chain.

\subsection{Comment on B$_{4/2}$ }

According to the Alaga rules, the ratio B$_{4/2}$ = B(E2;$4^+_1 \rightarrow 2^+_1$)/B(E2;$2^+_1 \rightarrow 0^+_\text{g.s.}$) is strictly larger than one. For an ideal rotor, B$_{4/2} = 10/7\sim1.43$, while for a harmonic vibrator, B$_{4/2}$ has an exact value of~2. In the current description of nuclear structure, B$_{4/2}$ can only have a value lower than 1 for structures conserving seniority (almost only found in semi-magic nuclei) and, in principle nuclei having shape-coexistence, but no example of the latter has been observed so far.

Cakirli and collaborators~\cite{Cakirli2004} carried out an extensive survey and found a few isolated cases with B$_{4/2}<1$ that could not be explained by either seniority or shape-coexistence. Subsequent experiments have re-measured with greater accuracy some of the most relevant of those isotopes and found important discrepancies for B(E2;$4^+_1 \rightarrow 2^+_1$) that made the B$_{4/2}$ values significantly larger than 1~\cite{Williams2006,Radeck2012,Zhu2017}.

Recent works~\cite{Cederwall2018,Esmaylzadeh2018} have suggested that the transitional neutron-deficient Hg isotopes could have B$_{4/2}$ values lower than 1. These suggestions arise from the $4^+_1$ half-life ($T_{1/2}$=7.2(3)~ps) quoted for $^{198}$Hg in the current edition of the Nuclear Data Sheets~\cite{Xialong2009,HUANG2016}, which in turn yields B$_{4/2}$=0.38(14). But the works cited in the compilation measured B(E2;$4^+_1 \rightarrow 2^+_1$) of 0.296(13)~$e^2b^2$~\cite{Bockisch1979} and 0.307(24)~$e^2b^2$~\cite{Gunther1981}, in perfect agreement, yielding $T_{1/2}$=1.80(8)~ps, which returns B$_{4/2}$=1.56(19). Moreover, this is the evaluated value from previous editions of the Nuclear Data Sheets~\cite{CHUNMEI2002}. To the best knowledge of the authors, no new work has been published that supports the 7.2(3)~ps half-life, and thus it should be replaced back in the compilations for the previous $T_{1/2}$=1.80(8)~ps one, value that is in agreement with the upper limit measured in this work. Likewise, all the B$_{4/2}$ values obtained from this work (see Tab.~\ref{tab:lifetimes}), are, within uncertainties, above 1. This includes the T$_{1/2}(4^+_1)$ upper limit of $^{194}$Hg measured in this work, which, as opposed to the results presented in Ref.~\cite{Esmaylzadeh2018}, seems to favour B$_{4/2}>1$. The results presented in this work, thus, negate the hypothetical deviation from the current model of nuclear structure, at least for this isotopic chain.


\subsection{Lifetime of $2_2^+$ in $^{188}$Hg \label{sec:188Hg-2_2-lifetime}}

The literature value for the lifetime of the $2_2^+$ state in $^{188}$Hg was previously measured to be 141(31)~ps by Joshi \textit{et al}~\cite{JOSHI1994}. This value cannot be reconciled with the one observed in this work of $T_{1/2}<20$~ps. They used a different variation of the advanced time-delayed method (Ref.~\cite{Mach1989}), described in Ref.~\cite{Joshi1993}. Their method relies on measuring the time difference between the x-ray created by the electron capture and the one created by the conversion electron, plus the detection of said conversion electron in a Si(Li) detector to select a specific decay branch. Since all x-rays from the same isotope have the same energy, that method does not have the ability to distinguish between delayed and anti-delayed coincidences. Thus, instead of measuring the centroid shift between the delayed and anti-delayed coincidences, that method assumes that an increase of the width of the time difference measured by the TAC is proportional to a lifetime between the two detected x-rays. Since the START signal is given by the x-ray of the electron capture decay, the measurement of a lifetime with that method is susceptible of contribution from higher-lying states that $\gamma$-cascade into the measured one.

The GCDM method described in Section~\ref{sec:data_analysis} and employed in this work involves coincidence gates on specific $\gamma$-rays, not x-rays. This grants it the unambiguous selectivity of measuring the time difference between feeding and decaying $\gamma$-rays of a specific level, thus measuring its lifetime without the contribution of other levels. Moreover, distinguishing the delayed and anti-delayed coincidences allows for a more precise measurement than just the variation of the time difference distribution width. For these reasons, the authors believe the GCDM method to be more reliable than the one described by Joshi \textit{et al.} and the half-life limit presented in this work to be more solid.

\subsection{$\rho^2$ value of the $0_2^+\rightarrow 0_1^+$ in $^{200}$Hg}

The new half life measurement of the first excited $0^+$ state in $^{200}$Hg allows the electric monopole transition strength of the transition to the ground state to be determined for the first time. The $\rho^2$(E0) value of this $0_2^+\rightarrow 0_1^+$ transition is calculated to be $0.02(1)\times 10^{-3}$ based on the new half life of 8(4)\,ps. This is a fairly small value which compares well to others known in the local region which have been reported by the evaluation of Kib\'{e}di~\cite{Kibedi2005}. For example the $0^+ \rightarrow 0^+$ transition in $^{188}_{76}$Os$_{112}$ has a $\rho^2$(E0) value of 0.011(4)\,milliunits, and the $^{194}_{78}$Pt$_{116}$ and  $^{196}_{78}$Pt$_{118}$ isotopes have values reported as $<$0.17 and 0.19(10)\,milliunits, respectively.

This new value in $^{200}$Hg is an excellent benchmark of the $E0$ strength in a mercury isotope that is away from the shape coexistence region around the neutron mid-shell of $N \approx 104$. The lighter Hg isotopes display significantly larger $\rho^2$ values where the nature of the first excited 0$^+$ state is significantly different and the energies much closer. This shape coexistence scenario is responsible for driving the large $E0$ strength.

This further supports the configuration mixing scenario discussed in earlier sections. Large $\rho^2$(E0) are indicative of strong mixing between $0^+$ states ~\cite{Kibedi2005}. Both sets of calculations, IBM-2~\cite{Nomura2013} and IBM-CM~\cite{Garcia-Ramos2014}, predicted strong mixing for the lighter Hg isotopes and negligible for the heavier ones (IBM-2 is able to accurately reproduce the B(E2) values for $^{194,196,198,200}$Hg without including any mixing). This is confirmed by small $\rho^2$(E0) measured in this work for $^{200}$Hg, where no mixing is expected, in contrast with the strong $\rho^2$(E0) observed near the \textit{N}=104 mid-shell~\cite{HEY11}, where the mixing is much stronger. 

\section{Conclusions}

Using the LaBr$_3$(Ce) detector array of the GRIFFIN spectrometer at the TRIUMF-ISAC facility, we have carried out a systematic study of the transitional even $A=188-200$ mercury isotopes. The present work focused on extracting lifetimes in the pico- to nanosecond range using the GCDM. A total of 33 lifetimes were measured, 10 of them for the first time. Overall, very good agreement was found between the new results and previous measurements, with a significant improvement in precision for many cases. 

This increased precision allowed for meaningful comparison with IBM-2 and IBM-CM calculations. There is an excellent agreement between the deduced B(E2;$2^+_1 \rightarrow 0^+_1$) values from this work and the IBM-CM calculation. The lifetimes of the $4^+_1$ states were too short for GCDM, resulting in large relative error bars that prevented comparison to the calculations.

Both IBM studies predicted shape coexistence  in light Hg isotopes up to $^{188}$Hg, with maybe a weak effect in $^{190}$Hg. The new, more precise results presented in this work seem to validate this hypothesis, confirming the minimal mixing between normal and intruder structures for $^{192-200}$Hg. The ongoing analysis of the conversion electrons and $\gamma-\gamma$ angular correlations data collected in this experiment will shed more light into the evolution of configuration mixing in the Hg isotopic chain.

\begin{acknowledgments}
We would like to thank the operations and beam delivery staff at TRIUMF for providing the radioactive beam. We are grateful to J.~E.~Garc\'ia-Ramos (IBM-CM~\cite{Garcia-Ramos2014}) and K.~Nomura (IBM-2~\cite{Nomura2013}) for providing the results of their calculations and for fruitful discussions. The GRIFFIN infrastructure has been funded jointly by the Canada Foundation for Innovation, the British Columbia Knowledge Development Fund (BCKDF), the Ontario Ministry of Research and Innovation (ON-MRI), TRIUMF and the University of Guelph. TRIUMF receives funding through a contribution agreement through the National Research Council Canada. C.E.S. acknowledges support from the Canada Research Chairs program. This work was supported by the Natural Sciences and Engineering Research Council of Canada.
\end{acknowledgments}

\bibliography{Hg_bibliography}

\end{document}